\documentclass[12pt,preprint]{aastex}

\slugcomment{Not to appear in Nonlearned J., 45.}

\shorttitle{Symbiotic stars burning hydrogen }
\shortauthors{Orio et al.}

\begin{document}


\title{Two SMC Symbiotic stars undergoing steady hydrogen burning}


\author{M. Orio}
\affil{INAF, Osservatorio Astronomico di Padova, vicolo Osservatorio, 5,
I-35122 Padova, Italy\\
and Department of Astronomy, 475 N. Charter Str. University of Wisconsin,
Madison WI 53706, USA}
\email{orio@astro.wisc.edu}

\author{A. Zezas}
\affil{Harvard-Smithsonian Center for Astrophysics, 60 Garden Street, Cambridge, MA 02138}

\author{U. Munari}
\affil{INAF - Osservatorio Astronomico di Padova, vicolo Osservatorio, 5,
I-35122 Padova, Italy }

\author{A. Siviero}
\affil{Dipartimento di Astronomia, Universit\`a di Padova, vicolo Osservatorio,2 
I-35122 Padova, Italy }

\author{E. Tepedelenlioglu}
\affil{Physics Department, University of Wisconsin,
Madison WI 53706, USA}



\begin{abstract}
Two symbiotic stars in the Small Magellanic Cloud
(SMC), Lin 358 and SMC 3, have been supersoft X-ray sources (SSS) 
for more than 10 years.  We fit atmospheric and nebular models to their X-ray, 
optical and UV spectra obtained at different epochs.
The X-ray spectra are extremely soft, 
 and appear to be emitted by the white dwarf atmosphere
 and not by the nebula like in some other symbiotics.  We find that 
the white dwarf of SMC 3, the hottest of the two sources, had 
an approximately
constant effective temperature T$_{\rm eff} \simeq$ 500,000 K in  
 1993-1994, in 2003 and 2006,
 without indications of decrease in 12 years. The bolometric
 luminosity of this system in March of 2003 was more than an order of
 magnitude lower than three years later, however the time of the observation
 is consistent with a partial eclipse of the white dwarf,
previously found in  ROSAT and optical observations.
The red giant wind must be asymmetric or very clumpy in SMC 3,
because
the filling factor of the nebula around the source is not
 higher than 0.1. 
The compact object in Lin 358 has been at T$_{\rm eff} \geq$180,000  K since 
1993 and there is partial evidence of a moderate increase.
Atmospheric fits for both objects are obtained with log(g)=9, which
 is appropriate for white dwarf masses $>$1.18 M$_\odot$. 
No nova-like outbursts of these systems
have been recorded in the last 50 years, despite continuous optical monitoring
of the SMC, and there are no indications of
 cooling of the white dwarf, expected after a thermonuclear flash.
 We suggest therefore that in both 
 systems hydrogen is burning steadily in a shell on the WD
 at the rate $\simeq 10^{-7}$ M$_\odot$ year$^{-1}$, 
sufficiently high to inhibit nova-type mass loss as required for
 type Ia supernovae progenitors.  

\end{abstract}


\section{Introduction}

 Symbiotic stars are characterized by the simultaneous
occurrence in an apparently single object of two temperature
 regimes, differing by a factor 30 or more. The spectrum
 consists of a late type (M-) giant absorption spectrum,
 high excitation emission lines and a blue continuum. Symbiotics are
 interacting binary systems, in which  a cool
 giant star transfers mass onto  hot companion
(a white dwarf, or more rarely a subdwarf or a neutron star).
 Because of mass loss from the giant there is often a common gaseous envelope.
In most symbiotics containing a white dwarf (hereafter WD),
 the fundamental power source is nuclear
 burning  on the surface of a white dwarf (see Sokoloski 2003).
 Reviews can be found in Kenyon (1986) and  Sokoloski
(2003). X-ray emission has been detected from
 several symbiotics, and attributed to both the symbiotic wind-nebula component
 and the hot WD atmosphere (e.g. M\"urset et al. 1997). The
first is usually ``hard'' (in the ROSAT range) and called ``$\beta$
type'', while some symbiotics show also, or only, supersoft ``$\alpha$ type''
 X-ray emission. This component 
is attributed to the WD atmosphere if it is very luminous.

 Nova-like thermonuclear flashes trigger the largest amplitude
 outbursts (up to 8 mag) observed in some symbiotics, with recurrence times of
tens of years. Kovetz
 \& Prialnik (1994) demonstrated that mass accretion onto the WD at
rates of about 10$^{-8}$
M$_\odot$ year$^{-1}$ explain these large outbursts. In other systems,
 smaller amplitude (1-2 mag) optical outburst,
sometimes coincident with collimated jets, are observed.
  For a system of this second group, AG Dra, 
coordinated optical, UV and X-ray observations indicate that
the compact object is burning hydrogen in a shell at
 the high rate of $\approx$ few 10$^{-8}$ M$_\odot$ year $^{-1}$
(Greiner et al. 1997).
 In this system,
the  WD radius expands periodically causing feedback
on the secondary and variation in the mass transfer rate $\dot m$.
This causes the nuclear burning rate to slow down, and there is
 a recurrent increase in optical and UV luminosity and decrease
in X-ray luminosity. In a few other symbiotic systems,
  the compact object is extremely luminous
(L$_{\rm bol} >$ 1000 L$_\odot$) and lower effective temperature limits
 of the order of 100,000 K were determined 
from UV observations (Munari \& Buson 1994).
Five symbiotic stars, for which there is no record 
 of nova outburst, S32, SMC-3, Lin 358,
Draco C-1 and RX J0550.0-7151, have even been observed to be 
{\it supersoft X-ray sources} (SSS)
 with effective temperatures as high as 200,000-500,000 K
 (e.g. Kahabka et al. 2004). RX J0550.0-7151  has recently been 
observed again with   {\sl Chandra} ACIS-S (PI Greiner), but it was  
 not detected again. However, Draco C-1, SMC 3 and Lin 358 were 
detected as supersoft X-ray sources at different epochs.  
Draco C-1 was observed twice with ROSAT (see M\"urset
 et al. 1997 and the ROSAT archive). In this paper we analyse
 and discuss recent Chandra and XMM-Newton observations
of Lin 358 and SMC 3.  We will show that these two objects  are still bright SSS
 after more than 10 years, at constant (or even increasing)
 atmospheric temperature.

\section{The two sources and the ROSAT observations}

SMC 3 is a symbiotic star of the SMC 
 in a region with  low gas content, but close to
the bar, therefore within a young stellar population.
The secondary is an oxygen rich  giant (M\"urset et al. 1997).
 The optical spectrum shows emission lines of Ne V and Fe VII and
Raman scattering  indicating very high temperatures (Munari
\& Zwitter 2002). Jordan et al. (1996) fit 
the IUE ultraviolet spectrum taken on November 1993 
and the ROSAT spectrum of 1992 simultaneously with an atmospheric
 white dwarf model, deriving 260,000 K as a lower limit
for the WD temperature.  Kahabka (2004) argued 
that since the X-ray luminosity decreased during  1992,
 this simultaneous fit was misleading.
 Lin 358 was  one of the first supersoft X-ray sources
 discovered with  ROSAT and was monitored in 8 observations during  
 more than 7 years, from 1990 until 1998. Over a time scales of
 several months, there was at first  
significant decrease of flux, then an increase 
 between 1993 and 1995, with a minimum in 1994 during which
 SMC 3 was not detected. Kahabka (2004) 
interpreted it this behavior  a  partial eclipse of the WD by  the giant
component, since also optical data support an orbital period
of 4-4.8 years.  This author 
 fitted a blackbody at temperature in the range 25-45 eV,
and  a carbon edge to the ``bright state'' ROSAT spectra,
and  detected a moderate trend of spectral hardening as the
 count rate was decreasing, but the fit
was not possible at minimum.  With some  assumptions,
a mass accretion rate
 (2.8-8.2)$\times 10^{-7}$ M$_\odot$ was derived,
 which places this system at the highest $\dot m$ ever observed.

\section{New X-ray observations of SMC 3}

\subsection{A serendipitous Chandra observation of SMC 3 in 2003}
 
SMC-3 was serendipitously observed with Chandra ACIS-S 15.8
 arcminutes off-axis, with the S-1 back-illuminated chip,
on 2003 February 28-March 3.
A review of the instrument can be found in
Garmire et al. (2003). The supersoft spectrum is shown in Fig.1. 
Table 1 summarizes the main results of all the X-ray observations.
We measured the count rate only above 0.2 keV, to avoid calibration
 problems around the carbon edge.
We report the count rate in the 0.2-1 keV range in Table 1,
0.0064$\pm$0.0005 counts s$^{-1}$. One reason for excluding
counts above 1 keV is that 
 all the other observations were done with XMM-Newton and EPIC,
which has  a larger effective area in the supersoft
range, in the EPIC observations of these objects the 1-10 keV count rate is
 negligible. We must note, however, that the count rate in the 
 whole 0.2-10 keV ACIS-S range is  about 10\% higher
than in the 0.2-1 keV range, although 
 with higher background, 0.0077$\pm$0.0010 counts s$^{-1}$.
The count rate in the 1-10 keV range is  
0.0013$\pm$0.0006 counts s$^{-1}$, barely  2 $\sigma$ above
 the background.  Trying to fit the whole spectrum in the 0.2-10 keV
 with only the WD atmospheric models, we obtain
 the same parameters indicated in Table 1,  but with reduced $\chi^2 \geq$1.2.
Adding a {\sl MEKAL} or other
 thermal bremsstrahlung component component in XSPEC we do obtain $\chi^2$=1,
but with a plasma temperature kT$\simeq$60 keV, which is completely
 outside the ACIS range. A power law yields a very shallow slope 
($\nu$=0.7). We mention these fits, but we concluded that
the data in the 1-10 keV range add so much background compared
 to the signal of the source,   that the results are not statistically
significant. 

As we mentioned in the introduction,
X-ray emission has been also detected from another
 source in WD symbiotic systems, namely from the nebula
and/or the circumstellar wind (so called $\beta$ type spectra 
 by M\"urset et al., 1997; see also recent results by
 Sokoloski et al. 2006 and Wheatley et al. 2006), therefore
the high energy emission is probably real and due to the surrounding
material, but not sufficiently luminous for spectral fits. 

The count rate in this observation
 is not consistent with the ROSAT one at maximum. 
 Using PIMMS with Kahabka's (2004) spectral parameters, 
or more rigorously (given the large off-axis angle of the observation),
by  fitting atmospheric models to the spectrum (see
 below) we find that the measured count rate 
 indicates a  factor  $\geq$30 lower flux than expected during
 the long ``high'' state identified by Kahabka (2004) for this source. 
 The ACIS-S count rate was at the level 
expected at the beginning or at the end of the supposed partial 
 eclipse, rather than in the ``unobscured'' ROSAT observations.
This Chandra observation was done almost 9 years 
 years after the supposed eclipse:  if the WD luminosity was
close to minimum around this time, the orbital period is in the
range 4--4.8 years  inferred by Kahabka (2004).

   Even if the S-1 chip is sensitive to energy as low as 0.1 keV
like the on-axis S-3 chip,
 the point spread function is significantly
 degraded. Nevertheless, the uncertainty on the position of
 the X-ray source has 
been definitely improved compared to the ROSAT one. The position 
obtained with the WAVDETECT source detection algorithm in the
Ciao Chandra data analysis software is 
$\alpha_{2000}$=00,48,20.2 , $\delta_{2000}$=-73,31,52.8
 and it differs by 
only 0.874 arcsec from the optical position.
The absolute astrometry of Chandra has an uncertainty of
 $\approx$0.6"  and the uncertainty on the source
 centroid is only $\approx$0.5", resulting in an overall
positional uncertainty  $\approx$0.8".
 The peak of the radial surface brightness profile  appears be 
double peaked and smeared over
$\simeq$5 pixels, corresponding to 2.5 arcsec, but no
 other known sources are so close to cause confusion. 
The ACIS response below 0.3~keV is poorly calibrated, and the effective area
below 1.0~keV has degraded because of the build-up of contaminating material on
the detector window. However, the degradation has been well characterized
 and we
were able to extract and fit a spectrum for this source. In order to take
advantage of the latest calibration we reprocessed the data with
 Ciao v3.3.0.1 and
CALDB v.3.2.3. This version of CALDB in particular includes the
 non uniform
sensitivity degradation close to the detector edge
 (particularly important in the
case of SMC 3),
 and improved effective area determination below 1.0~keV.  We extracted a
spectrum from a 30" aperture centered on the source.
 The background
was extracted close to, but completely outside the source, and at
approximately the same distance
from the detector edge, in order to have  about the
 same effective area as the source
spectrum.  A response matrix and an ancillary response file were created
 with the
{\textit{mkacisrmf}} and {\textit{mkarf}} Ciao tools, respectively.
  In order to use $\chi^2$ statistics we grouped the spectrum in order to have
  more than 15 counts per bin.
We then fit the spectrum with several models, to
 obtain all the available constraints on the WD
temperature and effective gravity.
In Staminorivic et al.'s (2000) maps, we find that SMC 3
is located in a region with N(H)$\approx$10$^{21}$ cm$^{-2}$,
 albeit with a large uncertainty. As Table 1 shows, a  blackbody model
does not yield a good fit, with a reduced $\chi^2$=1.6, 
 temperature 36 eV,  and
 a bolometric luminosity about 10$^{36}$ erg s$^{-1}$.
A simple blackbody is only a crude approximation of the WD
atmospheric spectrum; for hot WD overestimates the
bolometric luminosity by an order of magnitude (e.g. Heise et al. 1994,
Balman et al. 1998).
A good fit is obtained instead for this spectrum with 
Non-Local-Thermal-Equilibrium (NLTE) atmospheric
models developed for in-shell hydrogen burning white dwarfs.
We tested a set of models developed by
Hartmann \& Heise (1997; we also included updated models for novae 
described in Orio et al. 2003a), and other models developed by
 Werner et al. (2004), Rauch et al. (2005). Several of
 these models yield a good fit with $\chi^2$=1, with
 effective temperature in the range 38-45 eV (439,000-520,000 K) and
 log(g)$\geq$8.5. The ACIS-S spectral resolution 
 does not allow us to discriminate between different
 chemical compositions. In Table 1 we indicate the parameters
 obtained with Hartmann's model of the atmosphere
on top of a Ne-O-Mg core, developed
 for N LMC 1995 (Orio et al. 2003a),
 and with a model that is the other extreme in terms of abundances,
 the  metal poor (``halo WD'') model of
 Werner et al. (2004), which includes only the elements up to
 calcium. The ``halo'' model has been calculated also for high effective
 gravity, massive WD made of C-O or Ne-O-Mg, but the low abundances imply 
that in the atmosphere the core elements have not been significantly
mixed.  Even if 
 the chemical compositions are very different, both models give a reasonable
fit and one cannot be chosen over the other with the quality
of the data we have.
 
 It is 
 very interesting that atmospheric models with log(g)$<$8.5 do not yield
 good fits. If the white dwarf has T$_{\rm eff} \simeq$500,000 K,
 the effective gravity must also be high: the luminosity would
 be unreasonably large, orders of magnitude over
 the Eddington value, if the radius was not quite compact. 
  A fit  with a model of a Ne-O-Mg white dwarf
with log(g)=9, developed by Hartmann for N LMC 1995 (Orio et al.
2003a),  with $\chi^2$=1 and T$_{\rm eff}$=45
eV=522,000 K, is shown in Fig.1. The unabsorbed
flux, 10$^{-12}$ erg cm$^{-2}$ s$^{-1}$ in the 0.15-1 keV range,
 corresponds to a luminosity L${\rm x} = 4 \times 10^{35}$  erg s$^{-1}$
in the same range for a SMC distance 58 kpc (Sparke \& Gallagher 2000).
Given the uncertainties in the calibration, we
consider the spectral results only indicative. We also
 note that broad-band spectra allow a reasonable estimate
of the effective temperature, but it is not surprising that
 models with  
different chemical compositions fit the data, because
this distinction can
be done only with grating observations of the most luminous sources
(e.g. Rauch et al. 2003, Lanz et al. 2005). 

\subsection{A serendipitous XMM-Newton observation of SMC-3 in 2006}

 SMC-3 was observed again off-axis with
 XMM-Newton for 22.9 ksec on 2006 March 19,
 in a survey of the X-ray binaries population in the SMC
(P.I. A. Zezas).  A description of the satellite is 
found in Jansen et al. (2001). For
a description of EPIC-pn  we refer to Str\"uder et al. (2001),
and for the  EPIC-MOS to Turner et al. (2001). 
Both for this observation and for the
one of Lin 358 described below, we extracted and analysed the data using
the ESA
XMM Science Analysis System (SAS) software, version 7.0.0,
 and the latest calibration files available as of  2006 August.
We used only single events and the strictest screening criteria
(FLAG = 0), which are recommended procedures for the softest sources. 
No Optical Monitor data are available for this exposure.

 Unfortunately, the EPIC observation was badly effected
 by background flares. Only the first $\simeq$4200 seconds
 of pn and MOS observation and the final 2150 seconds of MOS
 exposures were usable. Because of the high
 background, the two MOS detectors were switched off during the observation
 and turned on again for 2150 seconds at the end of the pointing,
 while no more flares occurred, so two separate
 exposures are available for each MOS. SMC-3 was observed in an
 area of the MOS-2 detector with bad pixels
(rejected with the parameter FLAG=0), so we could only
 use the MOS-1 data.  Above 1 keV, the count rate is
only 1 $\sigma$ above the background level, and adding
 a thermal or power law component to the atmosphere does not improve any spectral fit. 
A summary of the observations and of the spectral fits is again given
in Table 1. The count rates in the 0.15-1 keV range
 were  1.115$\pm$0.020 cts s$^{-1}$ and
0.215$\pm$0.008 cts s$^{-1}$  with pn and
 MOS-1 in the first 4200 seconds,
 and 0.256$\pm$0.126 cts s$^{-1}$ in the last
 2150 seconds of MOS-1 observations. These count rates,
unlike the one of the Chandra observation, are consistent with
 the ROSAT count rate out of the supposed eclipse. 

 Already in the first 4200 seconds, both  
 pn and MOS-1 data already have a significantly
 high background in the soft range. We
 base our fits mainly on the final 
 2500 MOS-1 exposure, during which 
the background was back to normal level.  We try to use the 0.2-1 keV MOS
range to increase the signal of this very soft
 source, although the MOS calibration
 is satisfactory only from 0.3 keV.  We 
can fit the spectrum in the 0.2-1 keV range with models 
 with effective temperature about 500,000 K,
 and log(g)=9, with the a reduced $\chi^2$=1.
We find it significant and interesting that models
with log(g)$<$9 do not fit the spectra.
Sequences of white dwarf configurations with hot and massive
 white dwarfs were computed by Althaus et al. (2005) for various
 effective temperature and chemical compositions. Although
the calculations stop at T$_{\rm eff}$=150,000 K, they
clearly indicate that only white dwarf masses
M$_{\rm WD}>$ 1.18 M$_\odot$ reach log(g)$\geq$9 for  any temperature
 (see Fig. 7 and 8 of Althaus et al., 2005).
 The transition appears to occur around
1.18 M$_\odot$, which we can regard as the approximate lower limit
 for the WD Mass.

 The comparison with a blackbody fit is also shown 
in Table 1.  All models indicate 
approximately equal, or slightly higher temperature than 
 in the Chandra observation, but the luminosity appears to be
 greater by a factor of 45 with the O-Ne-Mg WD model,
or a factor of almost 100 with other models.
This is consistent with the occurrence 
of an eclipse in 2003. We note that the value of N(H) appears
to be unchanged or only slightly higher in 2003, which seems to
 be consistent with a real ``eclipse'' (caused by the red giant itself) than
 with a significant
contribution by an additional obscuration caused by red giant wind. 
Kahabka (2004) suggested an eclipse preceded and followed by obscuration
by wind material. However, our 
results are not consistent with an obscuration by 
a dense wind at the high mass loss rate (4.7 $\times$10$^{-6}$ M$_\odot$
 yr$^{-1}$) in Kahabka's model, which yields to an additional
 column density up to N(H)=10$^{21}$ cm$^{-2}$ at minimum.
 However, since 
 no nebular lines appear in the optical spectrum,
 we also rule out that the red giant wind is occurring at the rate 
 predicted by Kahabka's model, so we favor an eclipse without significant 
 wind obscuration. Our reasoning is based on the existing evidence in Galactic
 symbiotic systems: the high mass loss rate in the wind in Kahabka's
 model is typical of Mira stars, and an example is HM Sge (M\"urset
 et al. 1991), in which the optical spectrum is totally dominated by circumstellar 
emission of ionized gas with nebular lines like O III, N III, N I.
 On the other hand, Vogel \& Nussbaumer estimate that the red  component of AG Peg has
 a mass loss rate of $\simeq$10$^{-7}$ M$_\odot$ yr$^{-1}$, and no 
 nebular lines are detected in this system.
 Whether the time scale of the ROSAT
 dimming (about 6 months from beginning to minimum) can be consistent with  
 the lack of the additional wind obscuration is not quite clear,
 since several different, not well known parameters play a role.  
  In Figure 3 we show the best fit, with the 
model atmosphere of a metal poor WD with log(g)=9 (see Werner et al. 2004 and
 astro.uni-tuebingen.de/~rauch/in.html) obtained
 with N(H)=4.3$\times 10^{20}$ cm$^{-2}$, T$_{\rm eff}$=507,000 K 
and an unabsorbed luminosity 1.8 $\times 10^{-11}$ erg 
s$^{-1}$ in the 0.15-1 keV range.
 The ``noisy'' pn spectrum of the first 4200 seconds  can be fit in the broader
 0.15-1 keV range with almost the same parameters, but 
 obtained only a reduced $\chi^2 \simeq$1.2. 

\section{X-rays observations of Lin 358}

Lin 358 is another symbiotic system in a region of low N(H) in the SMC.
It was first detected in X-rays by Haberl et al. (2000).
 We re-examined the original exposure,
an archival serendipitous ROSAT-PSPC observation of Lin 358, done on 1992
October 2 for almost 65 kseconds. The background corrected count rate
 was 0.0189$\pm$0.0014 cts s$^{-1}$, and the spectrum is one of the softest
 ever observed with ROSAT. We found only
 an upper limit to the blackbody temperature, T$_{\rm bb}\leq$18  eV.

\subsection{New XMM-Newton observations of Lin 358}

Lin 358 was observed in a 30 ksec pointed observation 
with XMM-Newton, on 2003 November 16-17. The 
observation has been recently presented by Kahabka \& Haberl (2006)
although these authors only fit the data with a blackbody
model. We find that the observation yielded sufficiently high
 signal to noise to deserve a fit with detailed atmospheric models,
to determine the effective temperature 
and luminosity as reliably as possible. We analysed
the data fitting the same models used for SMC 3.

The position of the XMM-Newton source differs from the optical 
position of Lin 358 by 1.5 arcsec, which is within the 1 $\sigma$ 
uncertainty for EPIC MOS (see Kahaka et al., 2006).
Due to background flares, only $\simeq$15 ksec (half
 of the total exposure) are usable for
the analysis.  As shown in Table 1,
 the count rate measured with EPIC pn in the 0.15-1 keV range is
 0.1057$\pm$0.0033 cts s$^{-1}$.
The MOS-2 count rate is 0.0049$\pm$0.0006 keV in the 0.2-1 keV range,
and only 0.0013$\pm$0.0004 keV in the better calibrated 0.3-1 keV range.
This source is so soft that we used only the pn data for spectral fitting. 
As noted by Kahabka \& Haberl (2006), 
the blackbody model does not yield a good fit.
 The above authors proposed a way to improve
the fit by adding a hard, very low luminosity thermal bremsstrahlung component,
whose ``tail'' would modify the apparent spectrum of the 
luminous compact object.  However, this source is
one of the ``softest`` ever observed in X-rays and the count rate above 0.8 keV
is less than 1 $\sigma$ above the background, so we object 
that adding a component does not seem statistically and physically meaningful.

 The best fit with the atmospheric models is obtained with the
 models of Rauch et al.
 (2005), log(g)=9 and T=200,000 K, both with ``halo'' and ``solar''
 abundances (see Table 1), however it s not a perfect fit,
 since it yields $\chi^2$=1.2 in the 0.15-1 keV range.
The parameters of the fit shown in Table 1 and Fig. 2, are N(H)=8.9 $\times
10^{20}$ cm$^{-2}$, T$_{\rm eff}$=200,000 K (17 eV), and  an unabsorbed
luminosity 8.3 $\times 10^{38}$ erg cm$^{-2}$ s$^{-1}$ in the 0.15-1 keV range.
 only in this range,  clearly implying a super-Eddington bolometric luminosity.
The values of N(H) is higher than expected in this region 
of the SMC, which is very peripheral, although there are
 no H I maps covering this area.  Probably there is intrinsic absorption of
 circumstellar material due to the wind.
 There is a hint of some structure in Fig. 3,
 and we remind that wind emission lines in the supersoft
range have been observed for other supersoft X-ray sources:
Cal 87 (Greiner et al.
2003, Orio et al. 2003b), MR Vel (Motch et al.
 2002, Bearda et al. 2002), and novae V4743 Sgr (Ness et al. 2003) and
V382 Vel (Ness et al. 2005). 
By fitting the spectrum with only a WD atmospheric model, 
 we may be overestimating the luminosity if there are overimposed emission
lines. This may also explain the poor fit, but the quality
 of the data does not allow sophisticated modeling.

\section{A note on the short term variability in X-rays}

 Post-outburst novae appearing as supersoft X-ray sources
have been observed to undergo
significant variability on time scales of minutes and hours (e.g.
Orio et al. 2002, Drake et al. 2003, Leibowitz et al. 2006).
Although we can rule out ``obscurations'' during
the time span of these observations of the kind observed for
V4743 Sgr (Ness et al. 2003), we cannot rule out 
variability with amplitude $\leq$20\%,  because either the observation
 times or S/N, or both, are not sufficiently large.
 
\section{SMC 3 and Lin 358 in optical and ultraviolet, 10 years earlier: CLOUDY models}

The temperature of the compact object derived from 
the ROSAT data of SMC 3 has a large error bar (Kahabka 2004),
 and only an upper limit is obtained for Lin 358. 
To find additional information on the history of SMC 3 and Lin 358,
there is an additional component of the symbiotic system that
can be studied: the nebula, formed because  
 there is wind emission in most symbiotics. In both these symbiotics
 we detect bright emission lines, but as we mentioned above,
 the so called 
``nebular'' lines are absent, indicating that the density of the
 nebular material
at least at large distance from the red giant
 must be relatively low. This is consistent with the values of
 N(H) obtained in the X-ray fits. 
The material ejected by the red giant does not increase 
 the column density very significantly, especially
 for SMC 3.  We re-examined 
the high S/N ratio spectra of both SMC 3 and Lin 358,  in the 
3300-9100~\AA\ range and with a dispersion of 2.5~\AA/pixels,
taken at the 1.5 m telescope at ESO with the
Boller \& Chivens + CCD spectrograph on 1994 October 17.
 These spectra  were flux-calibrated and published
in the spectrophotometric atlas of symbiotic stars  of 
Munari and Zwitter (2002).
 The SMC-3 and Lin 358  spectra are Fig.6 and 8 of the
 Atlas, respectively. 
In Fig. 3 we show the optical spectrum of SMC 3, in which we
indicate the coronal lines of [Fe IX] and [Fe X], which
 are not marked in the Atlas but
are very important for the scopes of the present paper,
 because they clearly
 indicate a   photoionizing source at very high temperature.

 We assumed that the WD is the source of photoionization, and
 used a photoionization code
to model the spectra of the two symbiotics in order to estimate the radius
and temperature of the hot source, and the parameters of the
circumstellar nebular region. To make use of additional
observational constraints, we
have also fit the available archive IUE spectra of the
two stars (SWP~49298 for Lin 358, and SWP~47572 for SMC 3). There is
no report of major variability for Lin 358 in 1993-1994, so we 
combined the information of the optical and IUE spectra for the
 analysis even
if the data were obtained about one year earlier (1993 November 22).
For SMC-3, we note that the date of the UV IUE spectrum (April 30 1993) 
and of our optical one correspond to times of approximately equal 
 ROSAT count rates, although one observation falls in 
a period of decreasing X-ray luminosity, and the 
other one in a time of increasing X-ray luminosity. By combining the spectral
information, 
we assumed that the WD temperature of SMC-3 did not vary
and the partial obscuration observed with ROSAT was really
an eclipse (with or without wind obscuration).

The photoionization model we used is
CLOUDY\footnote{http://www.nublado.org/}.
Given the unknown intrinsic
properties of the nebulae and the absence of observational evidence of
winds, we tried assuming   as a first approximation in both systems  a
spherically symmetric, constant velocity wind in
 which the density is proportional to
r$^{-2}$ (r is  the distance from the ionizing source, which
is assumed to be  
 close to center of the nebula). We find that with
a r$^{-2}$ density profile, with and without multiple, concentric and
 distinct shells the line ratios are not consistent with those observed.
 A reasonable match  to the observed spectrum is obtained 
instead with a constant density profile. 
 High spatial resolution observations in radio and  optical imaging 
 of symbiotic stars has never shown spherical symmetry or
 a plain wind profile, but instead highly complex structures are detected 
 (e.g. Corradi 2001, Corradi et al. 2001) including jets, voids, shells,
 colliding winds, etc. The structure of these nebulae is so complicated, 
 that a constant density profile 
 is a first approximation as good any 
 analytical density law to explore the properties 
 of ionized gas at the order-of-magnitude level.
Assuming a r$^{-2}$ density law we also rely very critically on spherical
 symmetry, which in the case of these two symbiotics is very
 unlikely to be a good approximation. Most of the mass outflow is
expected to occur on the irradiated side of the red giant, since 
the white dwarf is so hot that irradiation effect are very important
(see discussion by   Munari \& Renzini 1992).
The results presented in Table 2 for a spherical, constant
density circumstellar nebula, are 
only as an exploratory tool of the actual nature of the objects.

To de-redden the optical and IUE spectra we assumed the
standard $R_{\rm V}$=3.1 extinction law from Fitzpatrick (1999) and an average  
SMC reddening $E_{\rm B-V}$=0.08 (Mateo, 1998).
The values obtained with CLOUDY are shown in Table 2.
R$_{\rm H II}$ is the radius inside which all hydrogen is ionized.
There usually is a rather abrupt transition to neutral hydrogen in the
nebula, within a distance 2 $\times 10^{-4}$ R$_{\rm H II}$.
R$_{\rm He II}$ is the radius inside which all helium is ionized,
(only a few percent smaller than R$_{\rm H II}$), and it is also the
 radius inside which the electron density n$_{\rm e}$ is
is calculated.

For SMC 3, we obtained a WD radius $R_{\rm WD}$=0.115~R$_\odot$
and a surface temperature $T_{\rm eff}$=5.05 $\times 10^5$~K.
This temperature is approximately the same obtained
in the later X-rays observations described above.
It is very well constrained by the 
coronal lines. However, at a SMC distance 58 kpc, the
 WD radius implies super-Eddington luminosity,
L$_{\rm bol} = 2.5 \times 10^{39}$ erg s$^{-1}$.
A smaller radius is derived raising the electronic 
density n$_{\rm e}$, but then the value of n$_{\rm e}$ would not be consistent
with observed ratios of
 fluxes in H$\alpha$, H$\beta$ and He II at 4686 \AA.  
Classical novae are observed at super-Eddington
luminosity for a limited length of time, up to few days,
 and a  theoretical
explanation involving a ``porous'' atmospheric structure has been
invoked (a discussion is found in Kato \& Hachisu, 2006)
 but a super-Eddington luminosity cannot last for years.
The only way to explain both 
 the WD temperature and the electron density necessary
 to reproduce the spectra, is that the nebula was not spherically
 symmetric, and had a filling factor of 0.1 at most.
This is, however,  very reasonable not only because the nebulae
of symbiotics have the complex geometry outlined above,
but also because the red giant wind is expected to be very clumpy
 (see   Crowley 2006, PhD Thesis at Trinity College, Dublin, Ireland). 

We note that a filling factor of 1 implies also an unrealistically
 large total mass of gas, 1.6 $\times 10^{-4}$ M$_\odot$. 
Since the nebular hydrogen is assumed to be
 totally ionized, from the model we cannot directly derive an intrinsic column
 of neutral hydrogen (with or without taking into account a filling factor) 
 to add to N(H) towards the SMC and compare to the X-ray spectral fits.
However, we note that the value of R(H II) obtained in the model
(18.8 AU or 2.8$\times 10^{14}$ cm, is more than 5 times larger than
the value assumed by Kahabka (2004) in calculating the value
 of additional N(H) during the obscuration (note also
that Kahabka assumed a r$^{-2}$ law for the
 density in the wind). 

Fitting the spectra of Lin 358, we obtained 
a WD radius $R_{\rm WD}$=0.127~R$_\odot$ and a surface
temperature $T_{\rm eff}$=1.8 $\times 10^5$~K, corresponding
 to a bolometric luminosity 10$^{38}$ erg s$^{-1}$
(around the Eddington value). 
The nebula turns out to have a
H density 2.5 $\times 10^9$ cm$^{-3}$. The radius of the H II 
region  and  the mass of ionized gas 
 are 5.3 AU and 6.5 $\times 10^{-6}$~M$_\odot$ respectively,
 much smaller than in the SMC 3 model.
 The main inconsistency with the X-ray results in this case
is the large WD radius, which would indicate low
 effective gravity, and is reasonable only
if the WD atmosphere shrunk considerably at the
time of X-ray observations. It is however
more likely that, also in this case, the filling factor was
significantly lower than 1 and the WD was not at Eddington luminosity.
As an independent check on the model, we note that
for a SMC distance of 58~kpc, this nebula would radiate
an  H$\alpha$ flux of 8.9 $ \times 10^{-13}$~erg~cm$^{-2}$~sec$^{-1}$,
 which is very close to the de-reddened value 8.2 $ \times
 10^{-13}$~erg~cm$^{-2}$~sec$^{-1}$
estimated by Munari and Zwitter (2002) with other methods in 
the spectrophotometric Atlas. 
Although it is difficult to estimate
 the error in our temperature determination for the XMM observations,
due to remaining calibration uncertainties in
the softest range of EPIC pn, the temperature
 indicated by the X-ray spectral fit
 is close to 200,000 K, so it is likely that the WD
temperature increased between 1993-1994 and 2003.

\section{Conclusions}

The broad band X-ray spectra are consistent with  WD at temperature $\simeq$ 45 eV 
 and 18-26 eV for SMC 3 and for Lin 358, respectively.
 We also re-examined optical and IUE spectra of these symbiotic systems,
 done in 1993-1994.
Modeling the spectra with Cloudy, we found that the temperature of the central 
 source was approximately the same indicated by the recent X-ray observations for SMC 3, and 
 probably even lower for Lin 358. 
In the case of SMC-3, the 2003 observation was done around
 the time of a predicted partial eclipse or
 obscuration of the WD. We find the source
 at lower luminosity level, but with the same effective temperature
 of the later observation of 2006, consistently
 with an eclipse. 
This is the first time the effective temperature of the central source 
 in symbiotic stars is estimated to be  so high  for a long
 time. However, there are mostly 
lower limits for the WD temperature in the literature,
and repeated observations were missing before this work. 

The XMM-Newton X-ray spectra cannot be fit with models with log(g)$<$9
for both sources, a fact that seems to point out
 at white dwarfs masses M$_{\rm WD} >$1.18 M$_\odot$.
 Although the spectral fit does not clearly indicate what
 is the best set of abundances and whether there is mixing
 with the inner core material, such a massive WD
 is unlikely to be the result of steady
 accretion onto a He WD (with maximum initial mass close
 to 0.5 M$_\odot$), steadily accreting and burning
 at a rate of 10$^{-7}$ M$_\odot$ yr${-1}$ for several million
 years without significant evolutionary changes. Yungelson
 et al. (1985) find that He WD in symbiotics do not
 undergo steady H burning for long enough to reach
 the Chandrasekhar mass, but they also predict that 40 \% of
 C-O WD in symbiotics should be burning H steadily. We note that  
the SMC is constantly monitored in the optical, 
  but SMC 3 and Lin 358 have showed no nova-like outburst.
 Since the  outbursts of symbiotics  usually
last for at least a year, it is very likely that
outbursts would have been detected if they occurred.  
The high WD effective temperatures and the very likely
absence of mass-ejecting outbursts indicate
that these systems are in the highest regime of 
mass accretion rate, 
$m \dot \simeq 10^{-7}$ M$_\odot$ year$^{-1}$. This is  a regime at which all
 the energy produced in the thermonuclear burning is efficiently irradiated
at near-Eddington luminosity (e.g. Fujimoto 1982, Kovetz \& Prialnik 1984).   
This hypothesis should 
be tested, ruling out other scenarios. In one evolutionary
path the average $\dot m$ is lower,  T$_{\rm eff}$ 
varies cyclically, causing variable $\dot m$ and hydrogen burning 
 rate like in AG Dra
(Greiner et al. 1997). In another possible evolutionary
scenario, recurrent, mild, non-degenerate
 thermonuclear flashes occur without mass ejection. We should
 be able to rule out such scenarios  and
 confirm a constant, high $\dot m$ with frequents ``snapshot''
 X-ray and optical exposures.
In this way we will be able to  confirm that they
 are viable type Ia SN candidates. 
The eclipse that explains 
 the ``dimming'' at apparently constant effective temperature
 of the WD in SMC 3 can also be verified with frequent 
and quasi-simultaneous optical and X-ray observations. 
Confirming a constant T$_{\rm eff}$ on a long
time scale
is the critical test to track the evolution of these systems.
  
\acknowledgments
MO acknowledges the support of a Chandra archival data grant to study
 supersoft X-ray sources.
AZ acknowledges support from NASA grant NNG06GE68G and NASA LTSA grant 
NAG5-13056.
This research has made use of data obtained through the High Energy
Astrophysics Science Archive Research Center Online Service,
provided by the NASA/Goddard Space Flight Center.

Facilities: \facility{1.5 m telescope at the European Southern Observatory}

\clearpage

\begin{figure}
\begin{center}
\includegraphics[width=10.5cm,angle=-90]{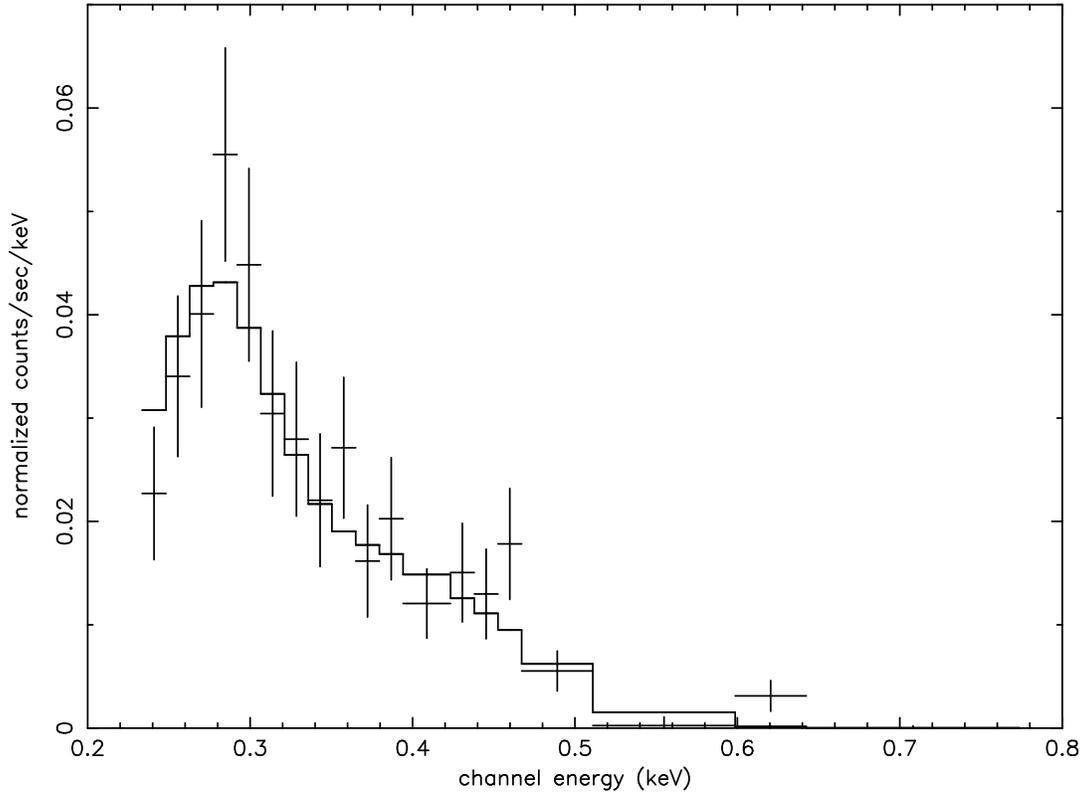}
\end{center}
\caption{The background corrected spectrum of SMC 3,
observed
 off-axis with Chandra ACIS-S, and the best fit with an atmospheric model with 
a Ne-O-Mg WD and T$_{\rm eff}$=45 eV.}
\end{figure}
\begin{figure}
\begin{center}
\includegraphics[width=11.5cm,angle=-90]{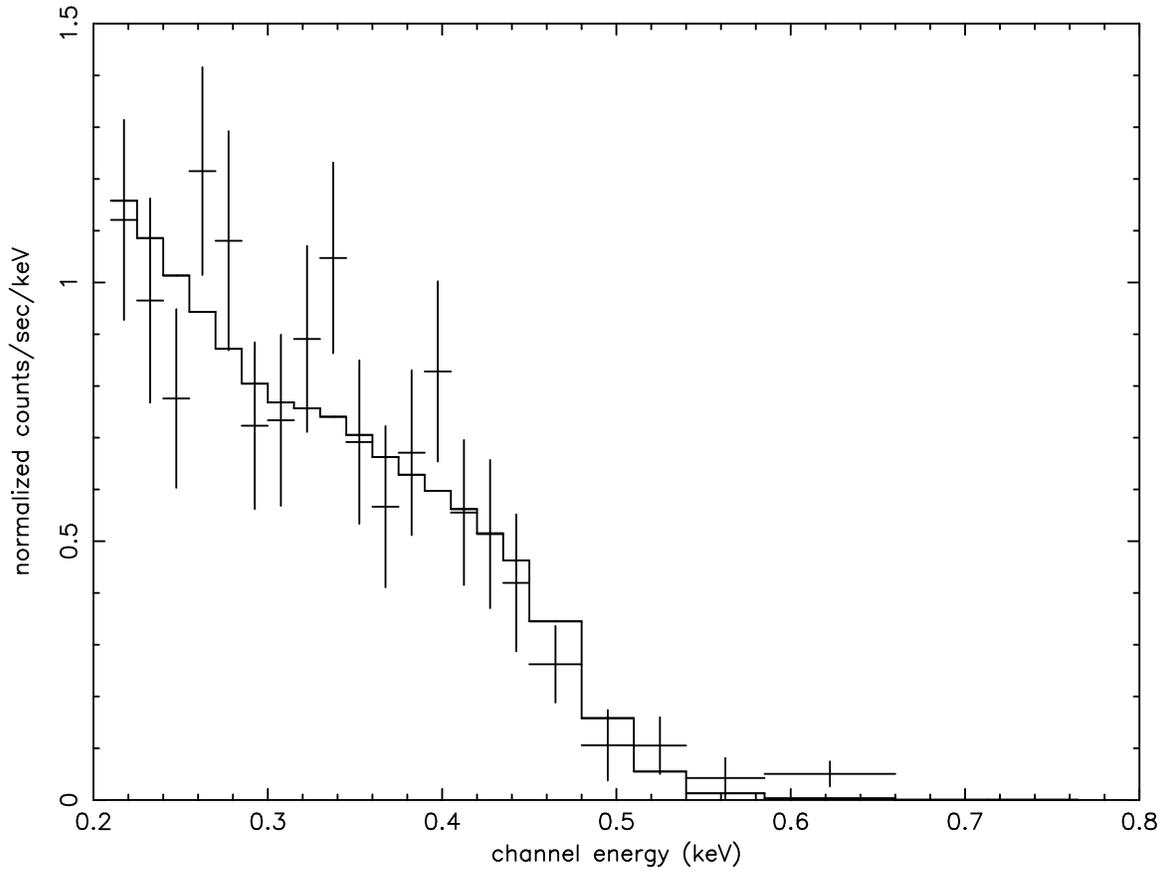}
\end{center}
\caption{The background corrected spectrum of SMC-3,
observed with EPIC-MOS1,  and the best fit with an atmospheric model with
a Ne-O-Mg WD and T$_{\rm eff}$=45 eV (see text).}
\end{figure}
\begin{figure}
\begin{center}
\includegraphics[width=11.5cm,angle=-90]{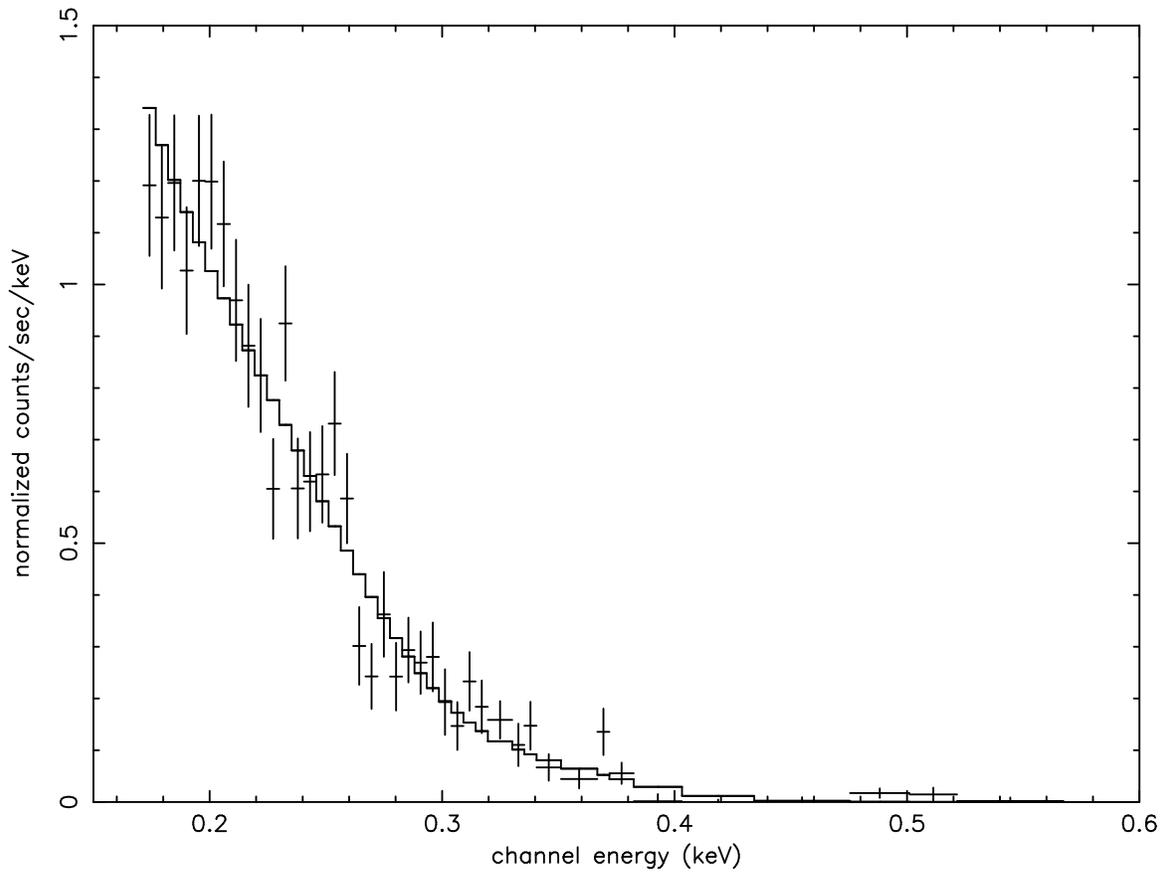}
\end{center}
\caption{The background corrected spectrum of Lin 358,
observed with EPIC-pn,  and the best fit with an atmospheric model
of a WD with
T$_{\rm eff}$=200,000 K (see text).}
\end{figure} 
\begin{figure}
\begin{center}
\includegraphics[width=9.5cm,angle=-90]{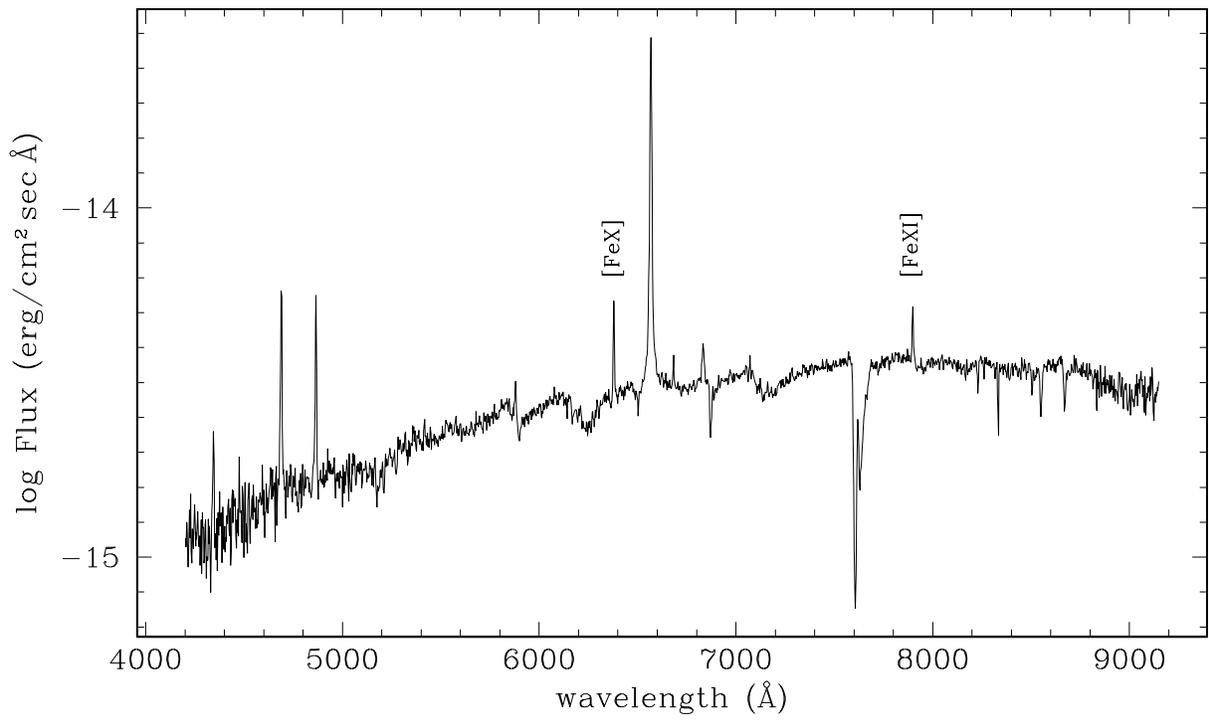}
\end{center}
\caption{The optical spectrum of SMC-3,
observed at ESO with the B\&C spectrograph
 at the 1.5m telescope. We indicate the coronal lines of [Fe IX] and [Fe X].}
\end{figure}
\newpage
\begin{deluxetable}{rrrrrrrrrr}
\tabletypesize{\scriptsize}
\rotate
\tablecolumns{11}
\tablewidth{0pc}
\tablecaption{{\small Count rates and spectral fit
parameters for the observations of SMC 3 and Lin 358 done in 2003 and 2006.
T is the blackbody temperature for the blackbody model, the effective temperature 
for the atmospheric models with
 log(g)=9. L$_{\rm x}$ is the unabsorbed luminosity in
the 0.15-1 keV range for the pn,
in the 0.2-1 keV range for EPIC MOS-1 and ACIS-S. 
The blackbody fit uses the bolometric luminosity
 as a parameter, and  although
the atmospheric models do not indicate the bolometric luminosity,
it is typically larger than L$_{\rm x}$ by a factor of almost 10.}}
\tablehead{ \colhead{Object} & \colhead{Instrument} &
\colhead{Date} & \colhead{count rate}  & 
\colhead{Model} &  \colhead{T (eV)} & \colhead{N(H) $\times 10^{20}$ cm$^{-2}$} &
\colhead{L$_{\rm x}$} & \colhead{L$_{\rm bol}$ (erg s$^{-1}$)} & \colhead{$\chi^2$/d.o.f} }
\startdata
SMC 3 & Chandra ACIS-S &  2003-2-28 & 0.0063$\pm$0.0005 & bb  & 36 & 7.5 & & 1.48 $\times 10^{36}$ & 1.6 \\
       &       &          &        & NeOMg      & 45 & 9.3 &  4  $\times 10^{35}$ & & 1.0 \\
       &       &          &        & Rauch(halo) & 39 & 4.0 & 9.1 $\times 10^{34}$ & & 1.0 \\
       &       &          &        &       &     &   &                       & & \\
SMC 3 & XMM MOS-1 &  2006-3-19 & 0.2174$\pm$0.0180   & bb  & 36 & 6.9
 &  & 1.3 $\times 10^{38}$ & 1.5 \\  
      &           &            &                   & NeOMg  & 45 & 8.4 & 1.8 $\times 10^{37}$
 & & 1.0 \\    
      &           &            &                   & Rauch(halo)  & 44 & 4.9 & 7.5 
 $\times 10^{36}$ & & 1.0 \\
       &       &          &        &  &    &    &                           & & \\
 SMC 3 & XMM pn  & 2006-3-19 & 1.1150$\pm$0.0244  & bb  & 36   & 8.0 &  & 2.2 $\times 10^{37}$ &
 1.3 \\
     &           &            &                   & NeOMg  & 47 & 3.1 &  10$^{37}$ 
 & & 1.2 \\
     &        &   &         & Rauch(halo) & 44 &  3.8 & 5.9 $\times 10^{36}$  & & 1.2 \\ 
       &       &          &        &           &     &        &              & & \\
 Lin 358 & XMM pn   & 2003-11-16 & 0.1057$\pm$0.0033  & bb   & 20 & 7.0 & & 2.5 $\times 10^{38}$  & 1.7 \\
         &    &   & & Rauch(sol)  & 17 & 8.5 & 8.4 $\times 10^{38}$ &  &  1.2 \\

\enddata
\end{deluxetable}
\newpage
\begin{deluxetable}{rrrrrr}
\rotate
\tablecolumns{6}
\tablewidth{0pc}
\tablecaption{{\small Nebular and WD
parameters (see explanation in the text) obtained using Cloudy models to
 reproduce the observed optical and UV spectra of SMC 3
 and Lin 358 in 1993-1994, in the approximation of
 spherical symmetry (and a filling factor of 1).}}
\tablehead{ \colhead{Object} & \colhead{T(WD) (eV)} & \colhead{R(WD) (R$_\odot$)}
& \colhead{M(H II) (M$_\odot$)} & \colhead{R (H II) (AU)} & 
\colhead{n$_{\rm e}$} } 
\startdata
SMC 3 & 43.5 & 0.115 & 1.6 $\times 10^{-4}$ & 18.8 & 2.05 $\times 10^9$ \\
Lin 358 & 15.5 & 0.127 & 6.5 $\times 10^{-6}$ & 5.3 & 2.50 $\times 10^9$\\ 
\enddata
\end{deluxetable}
%
%



\end{document}